\documentclass[english]{article}
\usepackage[english]{babel}

\usepackage{amsmath}
\usepackage{amsfonts}
\usepackage{cancel}
\usepackage[sort&compress]{natbib}
\usepackage{multirow}
\setcitestyle{numbers,square,comma}
\newcommand{\ha}{{\hat a}}
\newcommand{\hb}{{\hat b}}
\newcommand{\hc}{{\hat c}}

\newcommand{\haa}{{\hat \alpha}}
\newcommand{\hbb}{{\hat \beta}}
\newcommand{\hgg}{{\hat \gamma}}
\newcommand{\hdd}{{\hat \delta}}
\newcommand{\hee}{{\hat \epsilon}}

\def\cN{{\cal N}}
\def\no{\nonumber}

\begin{document}
\begin{center}

{\Large \bf Simplifying amplitudes in Maxwell-Einstein and
Yang-Mills-Einstein supergravities}

\bigskip

{\large Marco Chiodaroli}

\medskip

\small 

{ Max-Planck-Institut f\"ur Gravitationsphysik \\ 
             Albert-Einstein-Institut, Am M\"uhlenberg 1, 14476 Potsdam, Germany\\
             marco.chiodaroli@aei.mpg.de\\
}
\bigskip

\end{center}

\begin{abstract}
This article reviews recent progress in formulating  double-copy constructions 
 for scattering amplitudes in supergravity theories with $\cN=2$ supersymmetry in five and four spacetime dimensions. 
Particular attention is devoted to infinite families of Maxwell-Einstein theories with symmetric and homogeneous target spaces 
 and to Yang-Mills-Einstein theories with compact gauge groups. 
 Extension of the construction to theories with spontaneously-broken gauge symmetry is also discussed. 
\end{abstract}

\tableofcontents

%\maketitle

\section{Introduction and background}
\renewcommand{\theequation}{1.\arabic{equation}}
\setcounter{equation}{0}

Over the previous decade, scattering amplitudes in quantum field theories
involving gravity have been the object of renewed interest and intense investigation.
Calculations in the maximal and half-maximal supergravities have brought into focus 
simpler-than-expected structures and revealed improved ultraviolet (UV) behaviors.
While most of the explicit computations thus far have been within the purview of 
theories with a large number of supersymmetries, a growing research direction aims 
to extending this progress to theories with reduced supersymmetry. In this article,
I will discuss the extension of modern computational techniques, most prominently 
the double-copy construction, to infinite families of supergravities with eight supercharges.

The canonical work of Kawai, Lewellen and Tye (KLT) \citep{Kawai:1985xq} has first established that tree-level 
gauge-theory amplitudes are sufficient for constructing tree-level amplitudes in
the gravity theories which can be obtained 
from toroidal compactifications of string theory. 
The structure underlying the KLT relations has 
achieved a more modern formulation through the work of Bern, Carrasco and Johansson (BCJ), 
who expressed loop-level gravity 
amplitudes at the integrand level 
as ``double copies" of amplitudes in suitably-chosen gauge theories \cite{Bern:2008qj,Bern:2010ue}.
Their construction relies on the availability of gauge-theory amplitudes 
in which color and kinematic factors obey a duality known as color/kinematics (C/K) duality. 
This double-copy procedure
constitutes a breath-taking computational advance,
as it directly relates loop-level gravity amplitudes with gauge-theory amplitudes, which 
are significantly easier to obtain.
At the same time, 
the double copy has proven itself to be particularly well-suited for studying 
amplitudes in broader classes of theories with respect to its KLT precursor.  

A combination of the KLT property with unitarity-based methods \cite{Bern:1994zx}
and the BCJ double-copy construction have both been instrumental in conducting 
impressive multi-loop computations in the gravitational theories 
that result most amenable to perturbative calculations, that is theories which possess 
a large number of supersymmetries.
The theory with a maximal number of supersymmetries, $\cN=8$ supergravity,  
plays a key role in this context.  
This theory was first constructed by Cremmer and Julia and by de Wit and Nicolai  \cite{Cremmer:1978ds,Cremmer:1979up,deWit:1982bul}.
Explicit computations  
have shown that its UV behavior matches 
the one of the $\cN=4$ super-Yang-Mills (sYM) theory at least through four loops
\cite{Bern:2007hh, Bern:2008pv,Bern:2009kd,Bern:2012uf}.
These results 
lend support to the conjecture that the 
theory might be perturbatively UV-finite in four dimensions \cite{Bern:2006kd} and thus constitute 
the first known example 
of a mathematically-consistent quantum field theory of gravity.\footnote{\small Reviewing
arguments for and against finiteness  is beyond the scope of this note.
Ultimately, calculations at even higher loop order are necessary to establish conclusively the UV properties of the theory.}

Similar calculations have also been carried out  for the half-maximal theory --- $\cN=4$ supergravity \cite{Bern:2012cd,Bern:2012gh,Bern:2013qca,Bern:2013uka,Bern:2014lha}
--- and more recently for $\cN=5$ supergravity \cite{Bern:2014sna}.
In the absence of additional matter, the former theory appears to be finite at three loops in four dimensions and to diverge at four loops. 
The latter theory is UV-finite through at least four loops. At the moment, there are 
no widely-accepted symmetry arguments explaining the finiteness of pure $\cN=4$ supergravity at three loops and of $\cN=5$ supergravity at four loops.
The improved UV behavior of these theories is linked to the presence of enhanced cancellations 
between different terms of a diagrammatic presentation of the amplitude.
Interestingly, the four-loop divergence of pure $\cN=4$ supergravity appears to be related to a $U(1)$ 
quantum anomaly \cite{Bern:2013uka,Carrasco:2013ypa}. 
Additionally, recent calculations in gravity same-helicity amplitudes have shown that adding evanescent operators to the gravity action 
can alter the UV divergence of the theory while keeping the amplitude's dependence on the renormalization scale unchanged \cite{Bern:2015xsa}.
This result suggests 
that the latter quantity should be the one regarded as the physically relevant.

A better understanding of enhanced cancellations, 
the link between anomalies and divergences, and the role
of evanescent contributions 
will be critical in exploring the UV properties of $\cN=8$ supergravity and in 
determining its fate as a potential theory of quantum gravity, at least in the perturbative context.
In turn, achieving this understanding requires the ability of conducting perturbative calculations in more general gravity theories,
i.e. in theories that are not as special as maximal or half-maximal supergravity.
Unsurprisingly, one of the major
research directions established since the advent 
of the double-copy construction has been its extension and application to broader arrays of theories.
In particular, it is natural to ask: 
Is the double-copy structure a general property of gravitational interactions? 

Earlier studies on double copies with reduced ($\cN < 4$) 
supersymmetry have investigated very special theories, such as the ones that 
can be obtained as truncations of $\cN=8$ supergravity \cite{Carrasco:2012ca,Chiodaroli:2013upa, Damgaard:2012fb}, 
pure supergravities \cite{Johansson:2014zca}, and 
Einstein gravity coupled to a dilaton and an antisymmetric tensor \cite{Bern:2013yya}.  
This review focuses on extending the construction to infinite classes 
of supergravities with eight supercharges --- $\cN=2$ theories in four and five spacetime dimensions.
Unlike their more supersymmetric relatives, 
these theories are no longer completely specified by their matter content alone,
in the sense that two theories with the same spectra can have different interaction terms while still preserving 
$\cN=2$ supersymmetry. Moreover, many of these theories cannot be obtained taking  field-theory limits of
toroidal compactifications of string theory. 
%For technical reasons, it is convenient to consider theories which can be uplifted to 
%five spacetime dimensions.
The aim of this note is to provide a summary and a pedagogical introduction;
the reader should consult refs. \cite{Chiodaroli:2014xia,Chiodaroli:2015rdg,Chiodaroli:2015wal}
for a complete treatment. 
Particular attention will be devoted to discussing how some
physical features which can arise in supergravities with reduced supersymmetry, such 
as the introduction of non-abelian gauge interactions and the supergravity Higgs mechanism, 
are translated in the double-copy language.

\section{Maxwell-Einstein and Yang-Mills-Einstein supergravities\label{secMESGTs}}
\renewcommand{\theequation}{2.\arabic{equation}}
\setcounter{equation}{0}

Maxwell-Einstein theories with $\cN=2$ supersymmetry in five dimensions have been explicitly 
known since the early 80s due to the work of G\"{u}naydin, Sierra, and Townsend \cite{Gunaydin:1983bi,Gunaydin:1984ak,Gunaydin:1984nt,Gunaydin:1986fg}. 
These theories give the coupling of the gravity multiplet to $n$ matter vector multiplets. 
To fix the notation, we write their bosonic fields as
%\begin{equation}
$\big( h_{\mu \nu} , A^{0}_\mu \big)  \oplus  (A^x_\mu , \phi^x )$, where $x=1,2,\ldots, n$ and 
%\end{equation}
$A^0_\mu$ is the vector  in the gravity multiplet. Their  bosonic  Lagrangian is
\begin{equation}
e^{-1}\mathcal{L} \! = \!-\frac{1}{2}R- \frac{1}{4}
{\stackrel{\circ}{a}}_{IJ} F_{\mu\nu}^{I}F^{J\mu\nu} \!\! - \! \frac{1}{2}
g_{xy}\partial_{\mu}\phi^{x} \partial^{\mu}
\phi^{y} + \!  \frac{e^{-1}}{6\sqrt{6}}C_{IJK}
\varepsilon^{\mu\nu\rho\sigma\lambda}F_{\mu\nu}^{I}
F_{\rho\sigma}^{J}A_{\lambda}^{K} ,  \label{Lbossugra}
\end{equation}
where the $F_{\mu\nu}^{I}$ ($I=0,1,\ldots, n$) are abelian 
field strengths. We note that the  symmetric tensor $C_{IJK}$ 
that appears in the $F \wedge F \wedge A$ term needs to be constant to preserve gauge invariance. 
The matrices ${\stackrel{\circ}{a}}_{IJ}$ and $g_{xy}$ are functions of the physical scalars. 
The authors of ref. \cite{Gunaydin:1983bi} employ an Ansatz for supergravity Lagrangian and  supersymmetry transformations 
which depends on generic functions of the scalar fields 
and use  invariance of the Lagrangian under supersymmetry and closure of the supersymmetry algebra to
derive a set of algebraic and differential constraints.  
The most general solution to these constraints is found introducing an auxiliary ambient space with coordinates 
$\xi^I$  and defining a cubic polynomial $\mathcal{V}(\xi)$ in terms of the $C$-tensor,
\begin{equation}
\mathcal{V}(\xi)\equiv C_{IJK}\xi^{I} \xi^{J} \xi^{K}\, .
\end{equation}
The cubic polynomial is used to introduce the ambient space metric
\begin{equation}\label{aij}
a_{IJ}(\xi)\equiv -\frac{1}{3}\frac{\partial}{\partial \xi^{I}}
\frac{\partial}{\partial \xi^{J}} \ln \mathcal{V}(\xi)\, .
\end{equation}
Then, the matrices ${\stackrel{\circ}{a}}_{IJ}$ and $g_{xy}$, as well as the other quantities in the Lagrangian, 
have a geometrical interpretation:
\begin{itemize}
 \item[$\bullet$] The $n$-dimensional target space ${\cal M}_5$ 
 with coordinates $\phi^x$ is defined as the hypersurface with 
\begin{equation}
{\cal V} (h)=C_{IJK}h^{I}h^{J}h^{K}=1 \qquad , \qquad h^I = \sqrt{2 \over 3} \xi^I ;\label{prepotentialJordan}
\end{equation}
 \item[$\bullet$] The matrix ${\stackrel{\circ}{a}}_{IJ}(\phi)$ 
which appears in the kinetic-energy term for the vector fields is  the
restriction of the ambient-space metric to ${\cal M}_5$,
\begin{equation}
{\stackrel{\circ}{a}}_{IJ}(\phi)=a_{IJ}\big|_{{\cal V}(h)=1}\; ;
\end{equation}
\item[$\bullet$] The metric $g_{xy}(\phi)$ in the kinetic-energy term for the scalars is the induced metric on ${\cal M}_5$, 
\begin{equation} g_{xy}(\phi) = \frac{3}{2}\left. \frac{\partial \xi^I}{\partial \phi^x}
\frac{\partial \xi^J}{\partial \phi^y} a_{IJ} \right|_{ {\cal V}(h) = 1} \ . \end{equation}
\end{itemize}

The key result is that all quantities  in the Maxwell-Einstein Lagrangian can be expressed 
in terms of the $C$-tensor.  
Since the $C$-tensor can be obtained by inspecting three-point amplitudes,
$\cN=2$ Maxwell-Einstein theories in five dimensions are uniquely specified by their three-point interactions. 
This is in contrast to Maxwell-Einstein theories that only exist in four dimensions, for which supersymmetry is not as constraining.

Requiring positive-definiteness of the scalar and vector kinetic terms 
at a base-point imposes a constraint relating the base-point $\xi^I = c^I$ and the $C$-tensor,
\begin{equation}
c^I = \sqrt{2 \over 3} C_{IJK} c^J c^K \ .
\end{equation}
%Furthermore, the pull-back of the $C$-tensor on the target space, $T_{xyz}$ is related to the target-space Riemann tensor as
%\begin{equation}
%K_{xyzu} = {4 \over 3} \Big( g_{x[u} g_{z]y} + T_{x[u}^{ \ \ w} T_{z]yw} \Big) \ . 
%\end{equation}
Choosing the pull-back of the $C$-tensor to be covariantly-constant results in a locally-symmetric target space. 
In turn, $C$-tensors  with this property can be obtained using the 
theory of Jordan algebras and identifying
${\cal V}(\xi)$ as the norm 
of a euclidean Jordan algebra of degree three. 
This construction has permitted 
to obtain two classes of supergravities based on symmetric target spaces. 
The first, named generic Jordan family, is  based on  an infinite family 
of reducible Jordan algebras. 
The second is based on hermitian $3 \times 3$ matrices with entries in the four division algebras $\mathbb{R},\mathbb{C},\mathbb{H}$, and $\mathbb{O}$, 
and gives the so-called Magical Supergravities. 

We give here only the cubic polynomials relevant to the cases of interest. 
Theories belonging to the generic Jordan family have \cite{Gunaydin:1983bi}
\begin{equation} 
{\cal V}(\xi) = \sqrt{2}\big( \xi^0 (\xi^1)^2 - \xi^0 (\xi^i)^2 \big) \ , \qquad  i=2,3,\ldots, n  \ .
\end{equation}
Their target spaces in five and four dimensions are the symmetric spaces
\begin{equation}
{\cal M}_5 = {SO(n -1  , 1 ) \over SO(n -1) } \times {SO(1,1)}  \ , \qquad
{\cal M}_4 = {SO(n , 2 ) \over SO(n) \times SO(2)} \times {SU(1,1) \over U(1)}  \ .
\end{equation}

A more general class of theories is associated to homogeneous target spaces. 
Van Proeyen and de Wit \cite{deWit:1991nm} have shown that the requirement of a transitive group of target-space isometries
permits  to write the cubic polynomial as
\begin{equation} 
{\cal V}(\xi) = \sqrt{2}\big( \xi^0 (\xi^1)^2 - \xi^0 (\xi^i)^2 \big) + \xi^1 (\xi^\alpha)^2  + \tilde \Gamma^i_{\alpha \beta} \xi^i \xi^\alpha \xi^\beta  \ ,
\label{homogeneous}
\end{equation}
where $i,j = 2,3, \ldots, q+2$, $\alpha, \beta$ are indices with range $r$, and the total number of vector multiplets in $5D$ is 
$n = 2 + q + r$.
The matrices $\tilde \Gamma^i_{\alpha \beta}$ are symmetric gamma matrices and form a representation of the euclidean 
Clifford algebra ${\cal C}(q+1,0)$. The parameter $r$ is a multiple of the dimension of the irreducible representation 
of ${\cal C}(q+1,0)$, which is denoted as ${\cal D}_q$,
\begin{equation}
r(P, \dot P ,q) = {\cal D}_q (P+ \dot P) \ \ \text{if}   \ \ q = 0,4 \ (\text{mod } 8) \ ,  \qquad
r(P,q) = {\cal D}_q P   \ \   \text{otherwise} ,  %  \ \ q \neq 0,4 \ (\text{mod } 8) \ ,  \qquad
\end{equation}
where $P,\dot P$ are non-negative integers.
The parameter $\dot P$ is introduced for the values of $q$ such that there exist two inequivalent irreducible 
representations of ${\cal C}(q+1,0)$. 
The generic Jordan family corresponds to either
$q=1$ and $P$ arbitrary or to $P=0$ and $q$ arbitrary. The magical theories correspond to $P=1, \dot P=0$, and $q=1,2,4,8$ and, in four dimensions, 
have the target spaces
\begin{eqnarray} \mathcal{M}_4(J_3^{\mathbb{R}})= \frac{Sp(6,\mathbb{R})}{ U(3)}, &\qquad &  \mathcal{M}_4(J_3^{\mathbb{H}})=\frac{SO^*(12)}{
U(6)}, \no \\ 
\mathcal{M}_4(J_3^{\mathbb{C}})
= \frac{SU(3,3)}{S(U(3)\times U(3))},   &\qquad&  \mathcal{M}_4(J_3^{\mathbb{O}})= \frac{E_{7(-25)}}{ E_6\times U(1)}  \ . \end{eqnarray}

So far we have considered only supergravities of the Maxwell-Einstein type. 
However, it is possible to promote a subgroup of the isometry group and/or of the R-symmetry group to a non-abelian gauge group 
as discussed in refs. \cite{Gunaydin:1984ak,Gunaydin:1984nt}. 
Isometry transformations act linearly on the ambient-space coordinates as 
\begin{equation}
\delta_{\alpha} \xi^I =  (M_r)^I_{\ J} \xi^J   \alpha^r \ , \qquad [ M_r, M_s ] = f_{rs}^{\ \ \ t} M_t   \ ,
\end{equation}
where the matrices $M_r$ leave the $C$-tensor invariant. The gauging procedure is particularly simple when we consider compact isometry gaugings 
(i.e. we do not gauge part of the  R-symmetry group) and we further restrict to the  case in which the 
vector fields furnish the adjoint representation of the gauge group plus additional singlets 
(i.e. non-trivial representations other than the adjoint are not present).
This class of gaugings is obtained by introducing covariant
derivatives and field strengths in the Lagrangian in eq. (\ref{Lbossugra}),
\begin{eqnarray}
\mathcal{D}_\mu \phi^x &=& \partial_{\mu} \phi^x + g_s A^r_\mu K^x_r \ , \qquad K^x_r = - \sqrt{ 3 \over 2} f^{r I J} h_I h^{J x} \ . \\
\mathcal{F}^I_{\mu\nu} &=& 2 \partial_{[\mu} A^I_{\nu]} + g_s f^I_{\ JK} A^J_\mu A^K_\nu \ ,
\end{eqnarray}
where $g_s$ is the gauge coupling constant. The antisymmetric tensors   
$f^{IJK}$ are equal to the structure constants of the gauge group when the three indices $I,J,K$ 
assume values corresponding to the vectors promoted to gluons and  to zero otherwise. 
Additionally, we need to covariantize the $F\wedge F \wedge A$ 
term in eq. (\ref{Lbossugra}) and add a Yukawa-like term to the fermionic part of the Lagrangian. 
However, supersymmetry does not require  a non-trivial scalar potential, and hence 
the vacua of the theory are still of the Minkowski class. 

This procedure yields large 
classes of $\cN=2$ Yang-Mills-Einstein supergravities with compact gauge groups.
According to the choice of base-point, the non-abelian gauge symmetry 
can be unbroken or spontaneously-broken. This review 
focuses on Yang-Mills-Einstein theories of the generic Jordan family, for which
 the gauge group is a subgroup of  $SO(n)$. 
All gauge groups can be accommodated in this construction, provided that $n$ is taken to be large enough.  
The canonical choice of  base-point is,
\begin{equation}
c_{V_s}^I = \Big( {1 \over \sqrt{2}} , 1  , V_s , 0, 0 \Big)\, .
\label{base} 
\end{equation}
With $V_s=0$, the Yang-Mills-Einstein theory is in the unbroken-gauge phase.
When $V_s \neq 0$, 
the theory is on the Coulomb branch and contains 
massive vector multiplets transforming in matter (non-adjoint) 
representations of the unbroken gauge group. 
A $SO(n)$ transformation can be employed to bring 
all base-points to the form (\ref{base}). Since, in general, the theory 
will be invariant only under the subgroup of $SO(n)$ which is gauged, this transformation 
will change the form of the structure constants $f^{IJK}$. 

It is convenient to label the massive vectors with an index $\alpha$
running over all (not necessarily irreducible)
matter representations, so that the bosonic spectrum of the theory becomes 
\begin{equation}
 \Big( A^a_\mu , \phi^a \Big) \ \oplus \
  W_{\alpha \mu}  \ \oplus \
 \overline{W}^{\alpha}_{\mu}\ . 
\end{equation}
The fields $W_{\alpha \mu}$ and $\overline{W}^\alpha_\mu$ transform in conjugate representations. 
This choice corresponds to re-defining  the gauge-group generators that do not commute with $T^2$ so that 
$(T^\alpha)^\dagger= T_\alpha$ in a similar fashion to the treatment of
root generators in the Chevalley basis of a simple Lie group.
The mass matrix for the massive vector multiplets is proportional to the preferred $U(1)$ 
generator corresponding to the direction singled out by the non-zero $V_s$,
\begin{equation}  m^{\ \alpha}_{\beta} = i g_s {V_s \over \sqrt{1 - V_s^2} } (f^{2  \ \alpha}_{\ \beta} )  \ .  \end{equation}
The following sections will discuss how this plethora of theories can be obtained from the double copy of suitably-chosen gauge theories.
It is important to point out that several theories of interest have not (for now) been considered. These include theories with some
matter hypermultiplets, supergravities with R-symmetry gaugings, theories with non-compact gauge groups, five-dimensional supergravities 
with massive tensors in matter representations, and theories with spontaneously-broken supersymmetry. 

\renewcommand{\theequation}{3.\arabic{equation}}
\setcounter{equation}{0}

\section{Double-copy construction}

We start by considering $L$-loop, $n$-point amplitudes in a gauge theory with fields
in the adjoint representation and in a set of matter representations.
For very large classes of theories, such amplitudes can be expressed in a form based on a set of cubic graphs in which all external or internal lines
are labeled by a gauge-group representation,
\begin{equation}
   {\cal A}^{(L)}_{n} = i^{L-1} g^{n-2+2L} \sum_{i \in \text{cubic}}\, \int \frac{d^{LD}\ell}{(2\pi)^{LD}} \frac{1}{S_i} \frac{c_{i} n_i}{D_i} \, .
   \label{gaugepresentation}
 \end{equation}
$S_i$ are symmetry factors relevant for loop amplitudes.
The color factors $c_i$ are obtained by contracting the invariant 
tensors with indices in the representations carried by the lines joining at each vertex. 
 The denominators $D_i$ are given by the products of propagators associated to the internal lines of the graph.
Finally, the numerator factors $n_i$ are functions of loop and external momenta, 
as well as polarization vectors (or spinor-helicity brackets) and, when applicable, invariant 
tensors of the global-symmetry group.

Color/kinematics duality \cite{Bern:2008qj,Bern:2010ue} is the requirement that 
the color factors in eq. (\ref{gaugepresentation}) have the same algebraic properties as the numerator factors. 
In particular, due to the gauge-group Jacobi relations and to 
the representation-matrices commutation relations, there will be triplets of graphs whose color factors add to zero. 
In this case, an amplitude presentation is said to obey C/K duality manifestly 
if the corresponding numerator factors obey the same relations,
\begin{eqnarray}
      n_i - n_j = n_k & \quad \Leftrightarrow \quad & c_i - c_j=c_k \, . \no 
%      n_i \rightarrow -n_i  ~~~&\Leftrightarrow&~~~ c_i \rightarrow -c_i \,.
\label{duality}
\end{eqnarray}
This formulation  requires the gauge group to be kept general since  
avoiding a specific choice for the gauge group prevents the color 
factors from obeying extra identities aside from the ones stemming from the Jacobi and commutation relations.  

The existence of C/K-satisfying amplitude presentations has been proven 
at tree level for $\cN=4$ sYM \cite{Bohr:2009rd, Stieberger:2009hq, Mafra:2011kj} and, by extension, for some other theories that can be obtained directly 
from string theory,
such as pure sYM theories with various amount of supersymmetry.
While at loop level this property has a conjectural status, there is a strong and growing 
body of evidence in its favor, at least in particular theories.
For $\cN=4$ sYM, amplitude presentations which obey the duality manifestly have been constructed 
up to four loops at four points \cite{Bern:2009kd} and up to two loops at five points \cite{Carrasco:2011mn}.
General methods for constructing BCJ numerators at one loop have been investigated in refs. \cite{Bjerrum-Bohr:2013iza}.
 
This property has also been established for a variety of other theories, 
including self-dual Yang-Mills (YM) \cite{Monteiro:2011pc,Boels:2013bi}, theories with higher-dimension operators
\cite{Broedel:2012rc}, QCD (including supersymmetric versions) \cite{Johansson:2015oia, delaCruz:2015dpa}, non-supersymmetric YM theories 
with various matter \cite{Nohle:2013bfa},
the non-linear sigma model \cite{Chen:2013fya},   
the Bagger-Lambert-Gustavsson theory in three dimensions \cite{Bargheer:2012gv,Huang:2012wr,Huang:2013kca}, 
and several non-supersymmetric theories that we will later review.

Duality-satisfying structures arise naturally in string 
theory \cite{Stieberger:2009hq, Bohr:2009rd}.
Indeed, the point-particle limit of string theory can be used as a method to construct duality-satisfying numerators
\cite{Mafra:2011kj,Mafra:2012kh,Broedel:2013tta,Mafra:2014oia,Mafra:2014gja,Mafra:2015mja,He:2015wgf}.

Note that gauge-theory amplitudes will obey the Ward identities dictated by gauge invariance, that is, 
when a gluon polarization vector for the $i$-th external particle is changed 
as ${\cal \epsilon}_i \rightarrow {\cal \epsilon}_i + f(k,{\cal \epsilon}) k_i$, 
the corresponding amplitude will be unchanged. At tree level, 
this requires  the variations of the numerator factors to obey,
\begin{equation}
n_i \rightarrow  n_i + \Delta_i \ , \qquad    \sum_{i \in \text{cubic}}\,  \frac{c_{i} \Delta_i}{D_i} = 0 \ . \label{gaugevariation}
\end{equation}
This identity relies only on the algebraic relations between the color factors.

We now turn to the double-copy prescription for amplitudes in theories involving gravity and consider pairs or gauge theories 
with identical gauge groups and the same set of representations. A gravity 
asymptotic state is associated to each gauge-invariant bilinear build with a pair of states from the two gauge theories.
Some states in the gravity theory, most notably the physical polarizations of the graviton and the other states in the same multiplet,
will be associated to bilinears with adjoint gauge-theory states.
For theories involving  non-adjoint matter representations, there will be additional sectors corresponding 
to tensor products involving
conjugate matter representations. Due to the requirement of gauge invariance, bilinears made of one adjoint and one 
non-adjoint state will be disregarded.

Once the double-copy has been established at the level of the free theory, amplitudes in the interacting theory can be obtained 
replacing the color factors of the first theory with the numerator factors of the second, 
\begin{equation}
{\cal M}^{(L)}_{n} = i^{L-1}\;\!\Big(\frac{\kappa}{2}\Big)^{n-2+2L} \sum_{i \in \text{cubic}}\, \int \frac{d^{LD}\ell}{(2\pi)^{LD}} \frac{1}{S_i} \frac{n_i \tilde{n}_i}{D_i} \,
.
\label{DCformula}
\end{equation}
Gauge invariance under linearized diffeomorphisms  descends 
from the gauge invariance of the gauge theories entering the construction. 
The corresponding Ward identity dictates that replacing the graviton polarization 
tensor for the $i$-th external particle as 
${\cal \epsilon}^i_{\mu \nu} \rightarrow {\cal \epsilon}^i_{\mu \nu} + f(k,{\cal \epsilon}) k^i_\mu q_\nu + f(k,{\cal \epsilon}) k^i_\nu q_\mu$, with $q \cdot k^i =0$,
does not change the amplitude. 
However, this follows directly from eq. (\ref{gaugevariation})
since color factors are replaced with numerator factors that have the same algebraic properties,
as required by C/K duality. While both gauge theories need to obey C/K duality, 
the construction leads to a sensible theory as 
long as the numerators of one of the two theories obey the duality manifestly. 

When a gauge theory possesses a global symmetry, that symmetry will be inherited by
the  gravity theory from the double copy. However, the symmetry that is
manifest in the construction will be only a subgroup of 
the symmetry of the resulting gravitational theory. In particular, amplitudes are covariant 
only under a compact subgroup of the $U$-duality group, while 
the non-compact symmetries of the Lagrangian manifest themselves through the amplitudes'
vanishing soft-limits \cite{ArkaniHamed:2008gz,Brodel:2009hu}. An alternative approach 
to understanding the gauge-theory origin of gravitational symmetries 
requires a formulation the double copy at the level of off-shell linearized supermultiplets 
\cite{Anastasiou:2013hba,Anastasiou:2014qba,Anastasiou:2015vba}.   

At tree level, the double-copy construction is equivalent to the KLT relations for theories that can be seen as the 
field-theory limit of some toroidal compactification of string theory. 
However, the KLT relations do not apply to theories not belonging to this class or,
alternatively, when the gauge theories entering the construction have non-adjoint fields.

The double-copy property is a feature of other modern approaches 
to scattering amplitudes, including most notably the scattering equation formalism
\cite{Cachazo:2013gna,Cachazo:2013hca,Cachazo:2013iea}.
Beyond the realm of scattering amplitudes, a growing body of literature has identified double-copy structures in the context of classical solutions.
The reader should consult refs. \cite{Monteiro:2014cda,Luna:2015paa,Ridgway:2015fdl,Luna:2016due,White:2016jzc} for exciting developments in this direction.
Finally, a comprehensive review of gauge-gravity relations can be found in ref. \cite{Carrasco:2015iwa}.

\section{Amplitudes from the double copy}
\renewcommand{\theequation}{4.\arabic{equation}}
\setcounter{equation}{0}

\subsection{Generic Jordan family of Maxwell-Einstein supergravities\label{secGJDC}}

This section applies the double-copy prescription to  the specific theories we have introduced in sec. 
\ref{secMESGTs}. At tree level and for a modest number of external particles, it is still possible, although increasingly cumbersome, 
to obtain amplitudes with a conventional Feynman-rule computation. For technical reasons, it is easier to conduct computations in four
spacetime dimensions. Hence, we start by dimensionally reducing the Lagrangian (\ref{Lbossugra}). Denoting 
as $e^{2\sigma}$, $A^{-1}_\mu$, and $A^I$ the $g_{55}$ component of the metric, the graviphoton field from dimensional reduction, 
and the components of the vector fields along the fifth dimension,
the bosonic Lagrangian reduced to four dimensions is expressed as follows \cite{Gunaydin:2005bf}: 
\begin{eqnarray} && \!\!\!\!\!\! \!\!\!\!\!\! e^{-1} {\cal L}^{4D}  =  -{1\over 2} R - {1 \over 16 } e^{3 \sigma} F^{-1}_{\mu \nu} F^{-1 \mu \nu} - 
{3 \over 4} \partial_{\mu} \sigma \partial^\mu \sigma \no - 
{3\over 4}  \text{\it \aa}_{IJ} \partial_\mu h^I \partial^\mu h^J \no
\\
&& \!\! - \! {1\over 2} e^{- 2 \sigma} \text{\it \aa}_{IJ} \partial_\mu A^I \partial^\mu A^J \! -\! {1\over 4} e^{\sigma} \text{\it \aa}_{IJ} 
\Big(  F^I_{\mu \nu} \! - \! {1 \over \sqrt{2}} F^{-1}_{\mu \nu} A^I \Big) \! \Big(  F^{J\mu \nu} \! - \! {1 \over \sqrt{2}} F^{-1 \mu \nu} A^J \Big) 
 \no \\
&& \! + \! {e^{-1} \over 2 \sqrt{6}} C_{IJK} \epsilon^{\mu\nu\rho\sigma} \! \Big\{ F^I_{\mu \nu} F^J_{\rho \sigma} A^K  \! \! -\!
{1 \over \sqrt{2}} F^I_{\mu \nu} F^{-1}_{\rho \sigma} A^J A^K \! + \! {1 \over 6} F^{-1}_{\mu \nu} F^{-1}_{\rho \sigma} A^I A^J A^K \Big\}.\no \\
\end{eqnarray}
The real scalars from the five-dimensional Lagrangian pair with the components 
of the vectors along the fifth dimension to give complex scalars in four dimensions, 
\begin{equation}
z^I  = {1 \over \sqrt{2}} A^I + {\sqrt{3} \over 2} i e^\sigma h^I \ . 
\end{equation}
An alternative route to write the four-dimensional Lagrangian involves a symplectic formulation based 
on a prepotential which, in turn, depends on the $C$-tensor, as explained in ref. \cite{Freedman:2012zz}.

Focusing on the case of the generic Jordan family with cubic polynomial (\ref{prepotentialJordan}), 
we expand around the base-point (\ref{base}) and, following ref. \cite{Chiodaroli:2014xia}, 
take the following steps:
\begin{enumerate}
 \item Dualize the graviphoton field $A^{-1}_\mu$;
 \item Redefine the vector fields as 
 \begin{eqnarray} A^{-1}_\mu &\rightarrow& {1 \over 4} \big( A^{-1}_\mu - A^{0}_\mu - \sqrt{2} A^{1}_\mu \big) \ , \no \\
  A^{0}_\mu &\rightarrow& {1 \over 2} \big( - A^{-1}_\mu + A^{0}_\mu - \sqrt{2} A^{1}_\mu \big) \ , \no \\
  A^{1}_\mu &\rightarrow&  - {1 \over \sqrt{2}} \big( A^{-1}_\mu + A^{0}_\mu \big) \ ; \end{eqnarray}
\item Dualize the new $A^{1}_\mu$ field and redefine $z^1 \rightarrow - i z^1$.
\end{enumerate}
Amplitudes at three and four points can be straightforwardly obtained with a standard Feynman-rule computation. 
One of the advantages of working in four dimensions 
is the possibility of employing the spinor-helicity formalism \cite{Elvang:2013cua}. 
As an example, the vector-vector-scalar amplitude is
\begin{equation}
{\cal M}^{(0)}_3 \big( 1 A^0_- , 2 A^A_- ,3 \bar z^B   \big) = - {\kappa \over 2} \langle 12 \rangle^2 \delta^{AB}  \ . \label{ex1}
\end{equation}

The large symmetry group of the supergravities belonging to the generic Jordan family 
facilitates the identification of the correct gauge theories entering their
double-copy construction. First, 
one of the gauge theories  needs to provide the desired amount of supersymmetry.
Hence, a natural candidate is a 
pure $\cN=2$ sYM theory. As before, we write only the bosonic part of the Lagrangian in four dimensions,
\begin{equation}
{\cal L}^{{\cal N}=2}_{bos} \,=\, -\frac{1}{4} F_{\mu\nu}^{\hat A}F^{\hat A \mu\nu} 
+(\overline{D_\mu \phi})^{\hat A}(D^\mu\phi)^{\hat A}
+ \frac{g^2}{2}  f^{\hat A \hat B \hat C} \phi^{\hat B}{\bar \phi}^{\hat C}
 f^{\hat A \hat D \hat E} \phi^{ \hat D}{\bar \phi}^{\hat E} \ , \label{gauge1}
\end{equation}
where $\hat A, \hat B, \hat C$ are adjoint gauge indices and $\phi$ is a complex scalar.
The second gauge theory is the dimensional 
reduction of a $(4+n)$-dimensional pure YM theory,
\newcommand{\allphi}[1]{#1}
\begin{equation}
{\cal L}^{{\cal N}=0} \! =  \! - \frac{1}{4} F_{\mu\nu}^{\hat A}F^{\hat A \mu\nu} 
\!+ \! \frac{1}{2}(D_\mu\phi^{A})^{\hat A}(D^\mu\phi^{ B})^{\hat A}
 \! - \! \frac{g^2}{4} f^{\hat A \hat B \hat C } \phi^{\hat B B}\phi^{\hat C C} f^{\hat A \hat D \hat E} \phi^{\hat D B}
\phi^{\hat E C}\! , \label{gauge2}
\end{equation}
where $\phi^A$ are real scalars labeled by the global indices $A,B,C=1, \ldots, n$. This theory has a
manifest $SO(n)$ global symmetry.
In this case,  both gauge theories  are known to obey C/K duality at least at tree level. 
However,  in the more involved examples discussed in the next subsections,
C/K duality will pose non-trivial constraints.

At this point, it is straightforward to check that 
the three-point amplitudes from the double-copy formula (\ref{DCformula}) reproduce the ones from the Lagrangian, 
provided that we identify the supergravity states with the $(\cN=2)\otimes (\cN=0)$ bilinears as \cite{Chiodaroli:2014xia}: 
\begin{eqnarray}
A^{-1}_- =  \bar \phi \otimes A_- \ ,    & ~~~ & h_- = A_- \otimes A_- \ , \no \\
A^0_- = \phi \otimes A_-  \ ,  & ~~~ & i \bar z^0 = A_+ \otimes A_-  \ , \no \\ % \end{eqnarray} \begin{eqnarray}
A^A_- = A_- \otimes \phi^A  \ ,  & ~~~ &  i \bar z^A = \bar \phi \otimes \phi^A \ , %\qquad A=1,2,\ldots,n \ , 
\label{map1}
\end{eqnarray}
with similar expressions for the CPT-conjugate states. With this map, the amplitude (\ref{ex1}) is obtained as 
\begin{equation}
{\cal M}^{(0)}_3 \big( 1 A^0_- , 2 A^A_- , 3 \bar z^B   \big) = {\cal A}^{(0)}_3 \big( 1 \phi , 2 A_- ,  3 \bar \phi   \big) \Big|_{\cN=2}
\otimes {\cal A}^{(0)}_3 \big( 1 A_- , 2 \phi^A , 3 \phi^B   \big) \Big|_{\cN=0}  \ .  
\end{equation}
The purpose of the dualization and field redefinitions enacted after  
reducing the theory to four dimensions has been precisely to
``align" the physical states from the supergravity Lagrangian 
with the ones from the double copy.  
Since the 
three-point amplitudes from the double copy match the ones 
from the Lagrangian, the construction 
will give the amplitudes of the correct theory also at higher point, at least at tree level (provided that we are able to find
 amplitude presentations obeying C/K duality manifestly). As explained before,
this is a consequence of supersymmetry and of the existence of a five-dimensional uplift for the theories. 

From a gauge-theory perspective, there are several interaction terms we can envisage adding to the Lagrangians (\ref{gauge1}) and (\ref{gauge2}).
Indeed, the following subsections 
will discuss
simple deformations of the gauge theories which have interesting  interpretations from the vintage point of the resulting
supergravity.

\subsection{Isometry gaugings}

A very natural deformation of the non-supersymmetric Lagrangian (\ref{gauge2}) is the introduction of cubic scalar 
couplings of the form,
\begin{equation}
  \label{def1} \delta {\cal L}^{\cN=0} = {\lambda \over 3!} g f^{\hat A \hat B \hat C} F^{ABC} \phi^{\hat A A} \phi^{\hat B B} \phi^{\hat C C} \ . 
\end{equation}
Here $F^{ABC}$ are antisymmetric tensors with three global indices and $\lambda$ is a real parameter. 
It is important to verify that the theory still obeys C/K duality after adding the term
(\ref{def1}). 
At four point, the order-$\lambda^2$ part of the  four-scalar amplitude has expression 
 \begin{equation}
\begin{aligned}
&\mathcal{A}^{(0)}_4(1 \phi^{A_1},2 \phi^{A_2}, 3 \phi^{A_3} ,4 \phi^{A_4 })\Big|_{\lambda^2}
= g^2 \lambda^2\left(\frac{1}{s}F^{A_1  A_2  B }F^{ A_3 A_4  B}  
f^{\hat A_1\hat A_2 \hat B} f^{\hat A_3 \hat A_4 \hat B}+
\right.\\
&\left.\qquad
                        \frac{1}{u}F^{A_3  A_1 B}F^{A_2  A_4  B}  
                        f^{\hat A_3 \hat A_1\hat B} f^{\hat A_2 {\hat A}_4 \hat B}
                        +\frac{1}{t}F^{ A_2  A_3  B}F^{ A_1 A_4  B}  
                        f^{\hat A_2 \hat A_3 \hat B} f^{\hat A_1 {\hat A}_4 \hat B} \right).\\
\end{aligned}
\label{notinpairs}
\end{equation}
Requiring that this amplitude obeys the duality produces a non-trivial constraint on the $F$-tensors, which
 need to obey Jacobi relations.
We have verified up to six points that C/K duality does not pose additional constraints in the YM-scalar theory deformed by the term (\ref{def1}) \cite{Chiodaroli:2014xia}. 
 
The interaction term (\ref{def1}) produces a non-vanishing amplitude between three scalars. 
Taking the double copy of this amplitude with a 
three-gluon amplitude in the supersymmetric gauge theory produces a non-vanishing gravity amplitude between three vector fields,
\begin{equation}
{\cal M}^{(0)}_3 \big( 1 A^A_- , 2 A^B_- , 3  A^C_+   \big) =   
 - \Big({\kappa \over 2 }\Big) {\lambda \over \sqrt{2}} {\langle 12 \rangle^3 \over \langle 23 \rangle \langle 31 \rangle}   F^{ABC} \ .
\end{equation}
Hence, the interaction term (\ref{def1}) produces non-abelian gauge interactions in the supergravity Lagrangian \cite{Chiodaroli:2014xia}. 
Inspection of the amplitudes from the double-copy leads to the identification of $F^{ABC}$ and $\lambda$ with the supergravity gauge-group 
structure constants and coupling constant,
\begin{equation}
\lambda \Big({\kappa \over 2} \Big) {F}^{ABC} = 2 g_s {f}_{\text{sg}}^{ABC} \ , \qquad A,B,C = 1, \ldots , n \ . \label{Fmap} 
\end{equation}
It is interesting to note that, in this case, the double copy has promoted a global symmetry in one of the gauge theories to a 
local symmetry in the resulting supergravity.

This construction extends previous single-trace amplitude results from ref. \cite{Bern:1999bx}. 
Amplitudes in Yang-Mills-Einstein theories have also been recently 
investigated  from the point of view of scattering equations \cite{Cachazo:2014nsa,Cachazo:2014xea}, 
ambitwistor string constructions \cite{Casali:2015vta,Adamo:2015gia},
and string amplitudes \cite{Stieberger:2014cea,Stieberger:2015kia,Stieberger:2015qja,Stieberger:2016lng}.

\subsection{Spontaneously-broken theories}

Having formulated a double-copy construction for Yang-Mills-Einstein theories, it is paramount to probe its validity away from the 
unbroken-gauge phase, i.e. for $V_s \neq 0$. 
In this case, a sensible candidate for 
the first gauge theory is a spontaneously-broken version of the Lagrangian (\ref{gauge1}).
This is achieved by considering  a scalar vacuum expectation value of the form
\begin{equation}
\langle \phi^a\rangle = V t^0 \, \delta^{a0} \ ,
\label{vev} 
\end{equation}
with $V$ real. Reality of the expectation value is a consequence 
of the existence of a five-dimensional uplift for the spontaneously-broken theory. 
We then write the gauge-group generators as 
$\{ t^{\hat A} \} = \{ t^{\hat a} , t^{\hat \alpha} , t_{\hat \alpha} \}$, where $(t^{\hat a})^\dagger = t^{\hat a}$, $(t_{\hat \alpha})^\dagger= t^{\hat \alpha}$,
and the index $\hat a$ runs 
over all unbroken generators. Scalar and vector fields  of the resulting spontaneously-broken gauge theory split accordingly as  
$A^{\hat  A}_\mu = \big( A^{\hat  a}_\mu , W_{\mu \hat  \alpha}, \overline{W}^{ \hat { \alpha}}_\mu\, \big)$ and $
\phi^{ \hat  A } = \big( \phi^{ \hat  a } , \varphi_{ \hat  \alpha}, \overline{\varphi}^{ \hat {\alpha}} \big)$.
With this notation, the structure constants $f^{\hat A \hat B \hat C}$ 
of the gauge group before symmetry breaking yield the structure constants of the unbroken gauge group,
representation matrices for the massive vector multiplets, and Clebsh-Gordan coefficients 
entering couplings between three matter representations,
\begin{equation}
f^{\ha \hb \hc}=  - i {\rm Tr}([t^\ha, t^\hb] t^\hc) ,
~~~f^{\ha \ \haa}_{ \ \hbb}= - i {\rm Tr}([t^\ha, (t^\hbb)^\dagger] t^\haa)  ,
~~~f^{\haa \  \hgg}_{ \ \hbb}= - i {\rm Tr}([t^\haa, (t^\hbb)^\dagger] t^\hgg) .
\label{tensors}\end{equation}

The tensors (\ref{tensors}) obey color relations 
inherited from the Jacobi identities of the gauge group before symmetry breaking. 
Aside from the Jacobi relations for the structure constants of the unbroken gauge group and the 
commutation relations for the representation matrices, the Clebsh-Gordan coefficients obey two extra relations,
\begin{eqnarray}
f^{\haa \ \hgg}_{\ \hee} f^{\hee \ \hbb}_{\ \hdd}  -
f^{\haa \ \hbb}_{\ \hee} f^{\hee \ \hgg }_{\ \hdd}  &=& 
f^{\haa \ \hee}_{\ \hdd} \, f^{\hgg \ \hbb}_{\ \hee} \,, \no\\ 
\Big(f^{\hbb \ \hee}_{\ \hgg} f_{\hee \ \hdd}^{\ \haa}
+ f^{\haa \ \hee}_{\ \hdd} f_{\hee \ \hgg}^{\ \hbb} 
 + f^{\ha \ \hbb}_{\ \hgg} f^{\ha \ \haa}_{\ \hdd}\Big)- (\haa\leftrightarrow\hbb)
&=& f^{\haa \ \hbb}_{\ \hee} f_{\hdd \ \hgg}^{\ \hee}  \, . 
\label{extraf} 
\end{eqnarray}
These identities arise only for representations obtained from the symmetry breaking of a larger gauge group.
The seven-term identity can be thought of as a  set of
three-term identities since, for any assignment of external masses,  at most three terms can be non-zero. 
%These three-term identities will be the ones imposed on the numerator factors in a duality-satisfying amplitude presentation.

The mass spectrum of the theory is given by $m_{\haa}^{\ \hbb} = {i g V} f^{0 \ \hbb}_{\ \hat{\alpha}}$, which
can be taken in a block-diagonal form with blocks corresponding to different irreducible representations.
Massive fields can be further organized into representations labeled by the $U(1)$ charge associated to 
their mass. The number of such representations needs to be kept general
in order to cover all possible symmetry-breaking 
patterns.
Amplitudes in the simple supersymmetric theory discussed here 
can be obtained considering higher-dimensional amplitudes with massless fields and assigning 
compact momenta proportional to the masses  to the external particles, as done e.g. in refs. 
\cite{Alday:2009zm,Boels:2010mj,Craig:2011ws,Naculich:2014naa,Naculich:2015zha}.

The non-supersymmetric gauge theory entering the construction is an extension of the YM-scalar theory (\ref{gauge2}).
In particular, the theory has a set of complex scalars $\varphi_\alpha$ which have 
the same masses as the ones in the spontaneously-broken theory and transform in conjugate representations. 
We then write the most general cubic couplings involving three scalars, obtaining the Lagrangian:
\begin{eqnarray}
{\cal L}'\!\! &= \!\!\! &-\frac{1}{4}F_{\mu\nu}^{\ha}F^{\mu\nu \ha}+\frac{1}{2}(D_\mu\phi^{a})^{\ha} 
(D^\mu\phi^{a})^{\ha} +  (\overline{D_\mu \varphi^{ \alpha}})_\haa  (D^\mu \varphi_{\alpha})^\haa 
- (m^2)_{\alpha}^{\ \beta} \, \overline{\varphi}^{ \alpha}_\haa  \varphi_{\beta }^{\ \haa}  \no \\
&&  + 
\frac{1}{3!}g \lambda  F^{abc} f^{\ha \hb \hc} \phi^{a \ha}\phi^{b \hb}\phi^{c \hc}
+ g \lambda  \Delta^{ab}F^{a \ \beta}_{\ \alpha} f^{\ha \ \hbb}_{\ \hgg} \phi^{b \ha} \overline{\varphi}^{ \alpha }_{\ \hbb} \varphi_{ \beta}^{\ \hgg} 
\no \\
&& +\frac{1}{2}g\lambda  F^{\alpha \ \gamma}_{\ \beta} f_{\haa \  \hgg}^{\ \hbb}  \varphi_{\alpha}^{\ \haa} \overline{\varphi}^{ \beta}_{\hbb} \varphi_{ \gamma}^{\ \hgg} 
+\frac{1}{2}g\lambda  F_{\alpha \ \gamma}^{\ \beta} f^{\haa \ \hgg}_{\ \hbb}  \overline{\varphi}^{\alpha}_{\ \haa} \varphi_{ \beta}^{\ \hbb} \overline{\varphi}^{ \gamma}_{\ \hgg} 
+ {\cal L}_{\text{contact}} \ ,  % \\  
%&&\null - \frac{g^2}{4} f^{\ha \hb \he} f^{\he \hc \hd} \phi^{a \ha}\phi^{a \hc} \phi^{b \hb}\phi^{b \hd} 
%- g^2 f^{\ha \ \hgg}_{\ \haa} f^{\hb \ \hbb}_{\ \hgg} \phi^{a \ha}\phi^{a \hb} \overline{\varphi}^\alpha_{\ \hbb} \varphi_\alpha^{\ \haa}  \no \\
%&& \null - {g^2} f_{\ \hee}^{\haa \ \hbb } f^{\ \hee}_{\hgg \ \hdd} \overline{\varphi}^{\alpha}_{\ \haa} \varphi^{\ \hgg}_\alpha \overline{\varphi}^{\beta}_{\ \hbb} \varphi^{\ \hdd}_\beta + \frac{g^2}{2} f^{\he \ \haa}_{\ \hbb} f^{\he \ \hgg}_{\ \hdd} \overline{\varphi}^{\alpha}_{\ \haa}\varphi^{\ \hbb}_\alpha  \overline{\varphi}^{\beta}_{\ \hgg}
%\varphi^{\ \hdd}_\beta  \, ,
\label{YMscalarGlobalBrokentrunc}
\end{eqnarray}
where $\Delta^{ab}$ is a diagonal matrix and $F^{abc}$, $F^{a \ \beta}_{\ \alpha}$, $F^{\alpha \ \gamma}_{\ \beta}$ are tensors in the global indices.
Imposing the relations (\ref{extraf}) on the  numerators of amplitudes with four massive scalars 
fixes the contact terms in the Lagrangian (\ref{YMscalarGlobalBrokentrunc})
and produces the following constraints:
\begin{itemize}
 \item[$\bullet$] The $F$-tensors need to collectively obey the same algebraic relations 
 as the structure constants, representation matrices, and Clebsh-Gordan symbols  of the gauge group. 
 They can be interpreted as pieces of the structure constants of a larger group, 
 which is spontaneously broken down to a subgroup with structure constants $F^{abc}$. 
 The symmetry-breaking pattern of the global group needs to be the same 
 as the one of the gauge group;
\item[$\bullet$] The massive scalars need to assume a block-diagonal form such that 
\begin{equation}
2 V f^{0 \ \haa}_{\ \hbb} \varphi^{\ \hbb}_\alpha = \lambda \rho F^{0 \ \beta}_{\ \alpha} \varphi^{\ \haa}_\beta  \  , \label{block}
\end{equation}
where $\rho$ is a free parameter, 
i.e. for each block, the charge of the preferred $U(1)$ gauge generator is proportional to the charge of a preferred global $U(1)$ generator;
\item[$\bullet$] The diagonal matrices $\Delta$  are fixed to
$\Delta^{ab} = \delta^{ab} + (\sqrt{1+\rho^2} -1) \delta^{a0}\delta^{0b}$. \end{itemize}
The reader should consult ref. \cite{Chiodaroli:2015rdg} for a complete treatment.
As before, identifying  supergravity three-point 
amplitudes from the double-copy (\ref{DCformula}) with the ones from a Feynman-rule computation
leads to the $(\cN=2)\times(\cN=0)$
field map:
\begin{eqnarray}
A^{-1}_- =  \bar \phi \otimes A_- \ ,    & ~~~ & h_- = A_- \otimes A_- \ , \no \\
A^0_- = \phi \otimes A_-  \ ,  & ~~~ & i \bar z^0 = A_+ \otimes A_-  \ , \no \\ % \end{eqnarray} \begin{eqnarray}
A^a_- = A_- \otimes \phi^a  \ ,  & ~~~ &  i \bar z^a = \bar \phi \otimes \phi^a \ , \no \\
\varphi_\alpha  =  \varphi   \otimes \varphi_\alpha \  , & ~~~ &  W_\alpha =  W  \otimes \varphi_\alpha \ . 
\label{map2}
\end{eqnarray}
Similarly, the two free parameters $\lambda$ and $\rho$ in the non-supersymmetric gauge theory are related to the supergravity parameters as 
\begin{equation}
\Big({\kappa \over 2} \Big) \lambda F^{ABC} = 2 g_s f^{ABC}_{\text{sg}}  \ , \qquad \rho = {V_s \over \sqrt{1 - V_s^2}} \ . \label{parameter}  
\end{equation}
Note that the massless supergravity states 
are obtained as double copies of adjoint gauge-theory states, 
while the massive sector is obtained from the double copy of gauge-theory states in matter representations.
While this review considers supergravities with eight supercharges, analogous constructions can be set forth for spontaneously-broken 
theories with $\cN=4$ or no supersymmetry by adjusting the spontaneously-broken gauge-theory factor (see Table \ref{tabfact})

\begin{table}[t]
\centering
\begin{tabular}{l|c|c}
Gravity coupled to \cancel{YM} & Left gauge theory  & Right gauge theory  
%& \# states  
\\
\hline 
${\cal N}=4$ \cancel{YM}ESG & ${\cal N}=4$ S\cancel{YM}  & YM + $\cancel{\phi^3}$ 
%&   
\\
${\cal N}=2$ \cancel{YM}ESG (gen.Jordan) & ${\cal N}=2$ S\cancel{YM}  & YM + $\cancel{\phi^3}$ 
%&   
\\
${\cal N}=0$ $\cancel{\rm YM}_{\rm DR}$-E + dilaton + $B^{\mu\nu}$ &  $\cancel{\rm YM}_{\rm DR}$  & YM + $\cancel{\phi^3}$ 
%&   
\\

\end{tabular}
\caption{\small Double-copy constructions for spontaneously-broken Yang-Mills-Einstein supergravities with different 
amounts of supersymmetry \cite{Chiodaroli:2015rdg}. $\cancel{\rm YM}_{\rm DR}$ indicates a spontaneously-broken YM-scalar theory from dimensional reduction.}
\label{tabfact}
\end{table}

\subsection{Homogeneous supergravities}

The analysis of the supergravity Higgs mechanism has underlined the key role played by matter (non-adjoint) 
representations in extending the double copy to larger classes of supergravities. 
Along these lines, another option for modifying the construction of sec. \ref{secGJDC}
is to include in the non-supersymmetric gauge theory fermions transforming in a matter representation. 
Adding adjoint fermions without at the same time introducing extra supersymmetries 
is forbidden by C/K duality, as discussed in ref. \cite{Chiodaroli:2013upa}. 
To have a non-trivial result, we also need to add to the supersymmetric theory some fields transforming in the conjugate representation.
In this regard, the minimal set of fields that can be added is two fermions and two scalars, corresponding to a half-hypermultiplet. 
To fix notation, we denote the physical states in the supersymmetric gauge theory as
\begin{displaymath}
 \big( A^{\hat a}_+ , \psi^{\hat a}_+ , \phi^{\hat a} \big)_G \oplus \big( A^{\hat a}_- , \psi^{\hat a}_- , \bar \phi^{\hat a} \big)_G \oplus 
\big( \chi_+ , \varphi_1 , \varphi_2 , \chi_- \big)_R \ , \end{displaymath}
where $R$ and $G$ label the matter representation and the adjoint representation. For generic $R$,  
we also need to add the CPT-conjugate states. However, here it is convenient to consider the case in which 
the representation $R$ is pseudo-real, i.e. there exist a unitary matrix $V$ such that $V T^{\hat a} V^\dagger = - (T^{\hat a})^*$, $VV^*=-1$. 
In this case, the half-hypermultiplet alone is CPT-self-conjugate. 

The non-supersymmetric gauge theory entering the construction is
a YM-scalar theory with extra fermions in the representation $R$, 
\begin{eqnarray} {\cal L} \!\!&= \!\!& - {1 \over 4} F^{\hat a}_{\mu \nu} F^{\hat a \mu \nu} + {1 \over 2} (D_\mu \phi^a)^{\hat a} (D^\mu \phi^a)^{\hat a}  \no 
+{i \over 2}  \overline{\lambda}^\alpha  D_{\mu} \gamma^\mu \lambda_\alpha \ \no \\ && \!\!\! \!\!\! \null  + {g \over 2} 
\phi^{a\hat a} \Gamma^{a \ \beta}_{\alpha} \overline{\lambda}^\alpha \gamma_5 T^{\hat a} \lambda_\beta 
 - {g^2 \over 4} f^{\hat a \hat b \hat e} f^{\hat c \hat d \hat e} \phi^{a \hat a} \phi^{b \hat b} \phi^{a \hat c} \phi^{b \hat d}
 . \ \ \label{Lfermion} \end{eqnarray}
$\hat a, \hat b$ are adjoint indices of the gauge group and 
$\alpha, \beta=1,\ldots,r$ and $a,b=1,\ldots,q+2$ are global indices. 
Spacetime spinor indices and gauge-group indices for the representation $R$ are not explicitly displayed. As before,
we introduce unconstrained  matrices $\Gamma^{a \ \beta}_{\alpha}$ in the global indices and let C/K duality establish their algebraic properties. 

\begin{table}[t]
\centering
\begin{tabular}{c@{$\qquad$}cccc}
$q$ &  ${\cal D}_q$ & $4D$ fermions $r(q,P,\dot{P})$ & conditions & flavor group \\  
\hline
$-1$&  $1$ & $P$ &  R &   $SO(P)$   \\ 
$0$ & $1$ & $P \! + \! \dot P$ & RW   & $SO(P)\! \! \times \! SO(\dot{P})$  \\
$1$  &  $2$ & $2P$ & R & $SO(P)$  \\
$2$  & $4$ & $4P$ & R or W  & $U(P)$ \\
$3$ &  $8$ & $8P$ & PR  & $USp(2P)$  \\
$4$  & $8$ & $8P \! + \! 8\dot P$ & PRW   & $USp(2P)\! \! \times \! USp(2\dot{P})$  \\
$5$  & $16$ & $16P$ & PR  & $USp(2P)$  \\
$6$  & $16$ & $16P$ & R or W &   $U(P)$  \\
$k \!+\! 8$  & $16 \, {\cal D}_k$ &  $16 \, r(k,P,\dot{P}) $ & as for $k$ & as for $k$     \\
\end{tabular}

\medskip

\caption{\small Parameters in the double-copy construction for homogeneous supergravities \cite{Chiodaroli:2015wal}. 
The third column gives the number of $4D$ irreducible spinors
in the non-supersymmetric gauge theory,
which can obey a reality (R), pseudo-reality (PR) or Weyl (W) conditions.  \label{tab1}} 
\end{table}

Imposing C/K duality on the four-point amplitudes with two adjoint scalars and two matter fermions 
gives the constraint \cite{Chiodaroli:2015wal}:
\begin{equation}
 n_u - n_t = n_s  \quad  \rightarrow  \quad \{ \Gamma^a , \Gamma^b \} = 2 \delta^{ab} \ ,  
\end{equation}
 i.e. that the  matrices $\Gamma^a$ form a $(q+2)$-dimensional Clifford algebra. 
 In turn, thanks to this relation, the non-supersymmetric theory can be regarded 
 as the dimensional reduction of a YM + fermions theory in $D=(q+6)$ dimensions.
 The problem of charting all possible supergravities obtained with this construction
 is thus equivalent to listing  irreducible spinor representations in $D=(q+6)$ dimensions. 
 A parameter $P$ is equal to 
 the number of irreducible spinors introduced in the gauge theory. 
 When $P$ is greater than one,  an additional flavor symmetry will be manifest in the 
 non-supersymmetric gauge theory. 
 The only difference with the standard treatment of spinors in $D$ dimensions is the presence of the matrix $V$ acting on the gauge indices, which
 enters reality (R) and pseudo-reality (PR) conditions of the form \cite{Chiodaroli:2015wal}
\begin{equation} 
\overline{\lambda} =  \lambda^t  {\cal C}_4 C V \ , \qquad
{\rm R:} \ \ C =   {\cal C}_{q}   \ , 
 \quad {\rm PR:} \ \ C =   {\cal C}_{q}  \Omega \    , \label{RPRcond}
\end{equation}
where ${\cal C}_{q}$  and ${\cal C}_{4}$ are the global  and spacetime charge-conjugation matrices which obey  the relations
${\cal C}_q \Gamma^a {\cal C}_q^{-1} = - \zeta (\Gamma^a)^t$, ${\cal C}_4 \gamma^\mu {\cal C}_4^{-1} = - \zeta (\gamma^\mu)^t$, $\zeta = \pm1$.
$\Omega$ is an antisymmetric real matrix acting on the flavor indices. 
These conditions can be employed to obtain irreducible spinor representations with a $q$-by-$q$ analysis. Note that the
reality (R) condition is the combination
of a pseudo-reality condition on the gauge-group indices with a pseudo-Majorana condition on the spinor indices.
In the particular case of $q=0,4$ (mod $8$), there are two inequivalent irreducible spinors, and the parameters $P, \dot P$  count the number of each.
The analysis is summarized in Table \ref{tab1}. The total range of the global indices $\alpha,\beta$ is fixed to
$r = {\cal D}_q   P  $ or
$r = {\cal D}_q  ( P +  \dot P) $, where $4 {\cal D}_q$ is the dimension of the irreducible $SO(q+5,1)$ spinor. 
The number of vector multiplets in the four-dimensional supergravity theory obtained with the double-copy construction 
is equal to $(3 +  q + r)$.  
This construction reproduces the classification of homogeneous supergravities by de Wit and van Proeyen \cite{deWit:1991nm}.
As for the previous cases, 
three-point amplitudes from the double copy (\ref{DCformula}) are compared with the ones from 
the supergravity Lagrangian. The two sets of amplitudes agree provided 
that the field map 
\begin{eqnarray}
A^{-1}_- =  \bar \phi \otimes A_- \ ,    & ~~~ & h_- = A_- \otimes A_- \ , \no \\
A^0_- = \phi \otimes A_-  \ ,  & ~~~ & i \bar z^0 = A_+ \otimes A_-  \ , \no \\ % \end{eqnarray} \begin{eqnarray}
A^a_- = A_- \otimes \phi^a  \ ,  & ~~~ &  i \bar z^a = \bar \phi \otimes \phi^a \ , \no \\
A_{\alpha -} = \chi_-  \otimes (U \lambda_-)_{\alpha}  \ ,  & ~~~ & i \bar z_{\alpha} = \chi_+ \otimes  (U\lambda_-)_{\alpha}  \ , \quad
\label{map}
\end{eqnarray}
is employed. $U$ is a unitary matrix whose exact form depends on the choice of $\Gamma^1$. 
With an appropriate choice for $U$, 
some of the entries of the $\Gamma^a$ matrices in the Yukawa couplings of the non-supersymmetric gauge theory reproduce 
the real $\tilde \Gamma^i$ matrices in the cubic polynomial  (\ref{homogeneous}),
\begin{equation} ( U^t  \Gamma^a C^{-1}  U ) = \big( -\mathbf{1}  ,    i \tilde \Gamma^i  \big) \ .
\end{equation}
It should be noted that the four magical supergravities can be recovered as particular cases of this construction. 
Finally, this framework can be modified to include supergravities with hypermultiplets by 
introducing in the non-supersymmetric gauge theory scalars transforming in the representation $R$.

\begin{table}
\centering
\begin{tabular}{l|c} 
\parbox{0.45\textwidth}{\centering \bf Gauge theories} &  {\bf Supergravity} \\[3pt]
\hline \small
{\bf GT1:} {\rm Pure $\cN=2$ sYM theory} & \small  {\rm Generic Jordan family of}  \\ \small
{\bf GT2:} {\rm YM+scalar theory from dim. red.} & \small {\rm Maxwell-Einstein supergravities} \\[3pt]
\hline \small
 {\bf GT1}: {\rm As before} & \small  \rm Yang-Mills-Einstein theories   \\ \small
{\bf GT2}: {\rm Add trilinear scalar couplings} & \small \rm (compact gaugings)\\[3pt]
\hline \small
{\bf GT1}: {\rm Spont. broken $\cN=2$ sYM theory} & \small \multirow{2}{*}{\rm Higgsed supergravities} \\ \small
{\bf GT2}: \rm Add massive scalars \\ 
\hline \small
{\bf GT1}: \rm Add hypers in representation $R$ \\ \small
{\bf GT2}: \rm Add fermions in representation $R$ & \small \rm Homogeneous supergravities \\ \small
\ \ \ \ \  \ \ \ \ \rm with Yukawa couplings \\[3pt] 
\hline \small
 {\bf GT1}: \rm Add hypers in representation $R$ & \small \rm \multirow{2}{*}{\rm Supergravities with hypermultiplets}\\ \small
{\bf GT2}: \rm Add scalars in representation $R$ \\
\end{tabular}
\medskip
\caption{\small Summary of double-copy constructions for supergravities with eight supercharges. The first row describes the basic construction, while rows 2-5 
list some variants. \label{tab2}} 
\end{table}

\section{Discussion and outlook}

This note has discussed  how the double-copy construction can be 
formulated in theories with $\cN=2$ supersymmetry in four and five dimensions. Our starting point has been an infinite family of 
supergravities with symmetric target spaces, the generic Jordan family of Maxwell-Einstein theories. 
In this case, two very simple gauge theories enter the construction: one is a pure sYM theory, the other a 
YM-scalar theory obtained by dimensional reduction. 
Simple modifications of the above theories produce intriguing effects in the resulting supergravity.
In particular, some physical features of supergravities with reduced supersymmetry, such as the possibility 
of gauging part of their isometry groups, are straightforwardly incorporated in the double-copy framework.
Results are summarized in Table~\ref{tab2}.

Extension of the double copy has relied on the possibility of incorporating non-adjoint representations in 
the gauge theories entering the construction.
In all cases, it has been possible to identify the supergravity given by the double-copy construction from its amplitudes at
three points. This is a consequence of supersymmetry combined with the existence 
of a five-dimensional uplift  for both the 
supergravity theory and the gauge theories entering the construction. 
Nevertheless, amplitudes at higher points have been considered as a  consistency check in refs. \cite{Chiodaroli:2014xia,Chiodaroli:2015rdg,Chiodaroli:2015wal}, 
where sample amplitudes at one loop have also been displayed. 

These developments open the door to loop-level computations in large families of supergravities with eight supercharges. As UV divergences 
are expected already at one loop for generic theories with matter, results will likely 
cast some insight into the abundance and role of enhanced cancellations and into 
the connection between higher-loop divergences and lower-loop quantum anomalies.
A systematic study of loop-level amplitudes is currently ongoing.

The success in extending the double copy  strongly suggests that 
the construction has a significant role to play in computations for generic gravity theories with reduced (or no) supersymmetry.
In this respect, the extension of the construction to the supergravity Higgs mechanism is 
particularly relevant, as spontaneously-broken gauge symmetry is a general feature of Yang-Mills-Einstein theories,
which generically have non-compact gauge groups that are broken down to a compact subgroup. Finally,
the last extension discussed in this 
article is critically important as homogeneous supergravities now constitute the largest family 
of theories previously studied in the  supergravity literature for which a double-copy construction is explicitly known. 
Such family includes theories that cannot be obtained as a toroidal compactification of string theory, marking a substantial departure 
from the setting in which the KLT relations and the double-copy construction were first introduced.

\section*{Acknowledgements}
I am very grateful to Murat G\"{u}naydin, Henrik Johansson, and Radu Roiban for collaboration
on refs. \cite{Chiodaroli:2014xia,Chiodaroli:2015rdg,Chiodaroli:2015wal}, 
on which this review is based. My research is supported by the German Research Foundation (DFG) 
through the Collaborative Research Centre ``Space-time-matter" (SFB 647, teilgruppe C6).

\end{document}